\documentclass[aps,prl,twocolumn,groupedaddress]{revtex4}
\usepackage{graphicx} 
\usepackage{dcolumn}
\usepackage{bm}
\begin{document}

\title{Nuclear spin-lattice relaxation rate and nonmagnetic
pair-breaking effect in electron-doped
Pr$_{0.91}$LaCe$_{0.09}$CuO$_{4-y}$: Signature of highly anisotropic $s$-wave gap} 
\author{Guo-meng Zhao$^{1,2}$} 
\affiliation{$^{1}$Department of Physics and Astronomy, 
California State University, Los Angeles, CA 90032, USA~\\
$^{2}$Department of Physics, Faculty of Science, Ningbo
University, Ningbo, P. R. China}

\begin{abstract}
We numerically calculate the nuclear spin-lattice 
relaxation rate ($R_{s}$) in the superconducting state in 
terms of anisotropic $s$-wave gaps.
By taking into account electron-phonon coupling, our
calculated $R_{s}$ for a conventional $s$-wave superconductor,
indium, is in quantitative agreement with the 
experimental data with a clear Hebel-Slichter peak. In contrast, by using the
highly anisotropic $s$-wave gaps inferred from the magnetic penetration depth and 
scanning tunneling microscopy, our
calculated $R_{s}$ curves for electron-doped Pr$_{0.91}$LaCe$_{0.09}$CuO$_{4-y}$
show no Hebel-Slichter
peak, in agreement with the experimental data. Finally, 
the observed weak nonmagnetic pair-breaking effect provides unambiguous
evidence for a highly anisotropic $s$-wave gap in this underdoped cuprate.

\end{abstract}
\maketitle 

The identification of the intrinsic gap symmetry in cuprates is crucial to the understanding of 
the microscopic pairing mechanism of high-temperature
superconductivity, which remains elusive for over twenty years.  
The superconducting transition temperatures $T_{c}$'s of hole-doped cuprates 
appear to be too high to be explained by the conventional phonon-mediated pairing
mechanism.  In contrast, the highest $T_{c}$ in electron-doped ($n$-type) 
cuprates is about 40 K, which is within the $T_{c}$ limit of the
conventional phonon-mediated mechanism. Indeed, earlier
\cite{Huang} and recent \cite{Zhao09} tunneling spectra in electron-doped cuprates show strong electron-phonon coupling
features, similar to the conventional superconductors. The
predominantly phonon-mediated pairing 
should be compatible with an $s$-wave gap. Many independent experiments designed to test the gap symmetry in the electron-doped system have led to controversial
conclusions. Surface-sensitive angle-resolved
photoemission spectroscopy (ARPES) \cite{Arm01,Matsui} implies a
$d$-wave gap with a maximum gap size of about 2.5 meV. This gap size
would imply a $T_{c}$ of about 14 K at the top surface, which is a factor of 1.9 lower than
the bulk $T_{c}$ of 26 K (Ref.~\cite{Matsui}). Surface and phase-sensitive
experiments \cite{Tsu} provide evidence for pure
$d$-wave order-parameter (OP) symmetry in optimally doped and overdoped $n$-type cuprates.
In contrast, nearly bulk-sensitive 
point-contact tunneling spectra along the CuO$_{2}$ planes \cite{Pon} show no zero-bias conductance 
peak (ZBCP) in optimally doped and overdoped samples
\cite{Kas,Bis,Qaz,Shan05,Shan08}, which argues against 
$d$-wave gap symmetry. The bulk-sensitive Raman
scattering data of Nd$_{1.85}$Ce$_{0.15}$CuO$_{4-y}$
imply an anisotropic $s$-wave gap with
a minimum gap $\Delta_{min}$ of 3.2 meV (Ref.~\cite{Zhaopreprint}),
which is very close to $\Delta_{min}$ = 3 meV inferred from the
magnetic penetration depth data \cite{Alff}.  Bulk-sensitive thermal conductivity 
\cite{Rick} and specific heat \cite{Ham}
data seem to support $d$-wave gap symmetry \cite{Rick,Ham} while the
same data can be quantitatively explained by nodeless $s$-wave gap
symmetry \cite{Zhaothermal}. The absence of the ``Hebel-Slichter'' or ``coherence'' peak
below $T_{c}$
in the nuclear spin-lattice relaxation rate ($R_{s}$)
of a slightly underdoped $n$-type Pr$_{0.91}$LaCe$_{0.09}$CuO$_{4-y}$ ($T_{c}$ =
24 K) \cite{Zheng} appears to argue against $s$-wave gap symmetry.  However, the absence of the coherence peak does not
necessarily rules out $s$-wave gap symmetry because there are several 
mechanisms that can suppress the coherence peak. One of the
mechanisms is quasi-particle damping due to strong electron-phonon
coupling \cite{Allen}.  Strong electron-electron correlation also 
leads to a strong suppression of 
the coherence peak \cite{Sca}. Furthermore, the coherence peak can 
be also reduced by gap anisotropy \cite{Hebel,Will}. Therefore,
the absence of the coherence peak in the $R_{s}$ data of Pr$_{0.91}$LaCe$_{0.09}$CuO$_{4-y}$ 
may arise from the combination of the intermediate electron-phonon coupling 
constant ($\lambda$ $\simeq$ 1) \cite{Huang}, strong electron-electron correlation,
and a highly anisotropic $s$-wave gap.

Here we present numerical
calculations of the nuclear spin-lattice relaxation rate in 
the superconducting state in terms of anisotropic $s$-wave gaps.
By taking into account electron-phonon coupling, our
calculated $R_{s}$ for a conventional $s$-wave superconductor,
indium, is in quantitative agreement with the 
experimental data with a clear Hebel-Slichter peak. In contrast, by using the
highly anisotropic $s$-wave gaps inferred from the magnetic penetration depth and 
scanning tunneling microscopy, our
calculated $R_{s}$ curves for Pr$_{0.91}$LaCe$_{0.09}$CuO$_{4-y}$ show no Hebel-Slichter
peak, in agreement with the experimental data. Finally, 
the observed weak nonmagnetic pair-breaking effect provides unambiguous
evidence for a highly anisotropic $s$-wave gap in this underdoped $n$-type cuprate.

The expression for the ratio of $R_{s}/R_{n}$  of an anisotropic superconductor
with a complex gap function is given by \cite{Will}

\begin{eqnarray}
\frac{R_{s}}{R_{n}}
=\frac{2}{k_{B}T}\int_{0}^{\infty}[<N(T, \omega)>^{2}+<M(T,
\omega)>^{2}]\nonumber \\
f(\omega)[1-f(\omega)]d\omega,
\end{eqnarray}

where $k_{B}$ is the Boltzmann constant, $f(\omega)$ is the Fermi-Dirac 
distribution function, $<N(T, \omega)>$ and  $<M(T,
\omega)>$ are the respective Fermi-surface averages of $N(T,
\omega)$ and $M(T,
\omega)$, which are given by 
\begin{equation}
N(T, \omega) = Re[\frac{\omega}{\sqrt{\omega^{2} -
\Delta^{2}(T, \vec{\Omega})}}],
\end{equation}

and 

\begin{equation}
M(T, \omega) = Re[\frac{\Delta (T, \vec{\Omega})}{\sqrt{\omega^{2} -
\Delta^{2}(T, \vec{\Omega})}}],
\end{equation}

where $\vec{\Omega}$ is the direction vector on the Fermi-surface. Taking into account electron-phonon coupling, $N(T,
\omega)$ and $M(T,
\omega)$ can be approximated by \cite{Will}

\begin{equation}
N(T, \omega) = Re[\frac{\omega}{\sqrt{\omega^{2} -
\Delta_{1}^{2}(T, \vec{\Omega})(1+i\delta)^{2}}}],
\end{equation}

and 

\begin{equation}
M(T, \omega) = Re[\frac{\Delta_{1}(T, \vec{\Omega})(1+i\delta)}{\sqrt{\omega^{2} -
\Delta_{1}^{2}(T, \vec{\Omega})(1+i\delta)^{2}}}],
\end{equation}

where $\Delta_{1}(T, \vec{\Omega})$ = $\Delta_{0} (T)\Delta
(\vec{\Omega})$ is the real part of the gap
function, $\Delta
(\vec{\Omega})$ determines the gap anisotropy, and $\delta$ is given by
\cite{Will}

\begin{eqnarray}
\delta = \frac{n\pi
\lambda}{1+\lambda}(\frac{T}{\theta_{D}})^{n}[\Gamma (n+\frac{5}{2})\zeta 
(n+\frac{3}{2})(\frac{k_{B}T}{2\Delta_{0}(T)})^{3/2}\nonumber\\
+\frac{\sqrt{\pi}}{n}(\frac{2\Delta_{0}(T)}{k_{B}T})^{3/2}\exp
(-\Delta_{0}(T)/k_{B}T)].
\end{eqnarray}

Here $\Gamma$ and $\zeta$ are the $\Gamma$ and Riemann $\zeta$ functions, 
respectively. At low temperatures, the electron-phonon spectral
function $\alpha{^2}(\omega)$$F(\omega)$ varies as $\omega^{n}$. In
the following numerical calculations for $R_{s}$, we
will adopt ``jellium'' model where $n$ = 2 (Ref.~\cite{Will}). 

\begin{figure}[htb]
     \vspace{-0.3cm}
    \includegraphics[height=6.2cm]{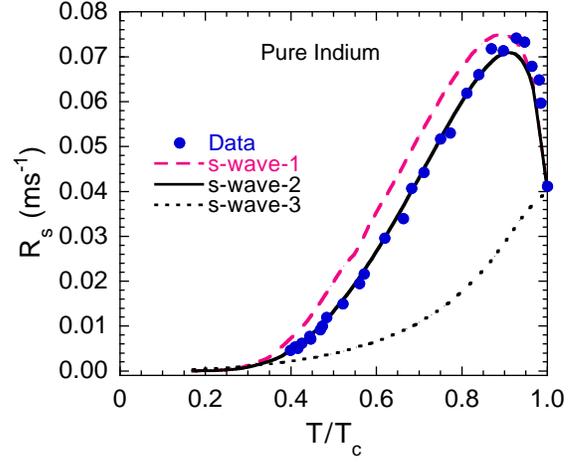}
     \vspace{-0.3cm}
 \caption[~]{Nuclear magnetic spin-lattice relaxation rate ($R_{s}$)
 of pure indium (solid circls). The data are taken from
 Ref.~\cite{Will}. The numerically calculated $R_{s}$ for
three gap functions: $\Delta
(\vec{\Omega})$ = 1 for isotropic $s$-wave gap ($s$-wave-1); $\Delta
(\vec{\Omega}) = (1 - 0.02\cos 4\phi)$ for a slightly anisotropic $s$-wave
gap ($s$-wave-2); $\Delta
(\vec{\Omega}) = (1 - 0.9\cos 4\phi)$ for a highly anisotropic $s$-wave
gap ($s$-wave-3). }
\end{figure}

We first present the results of numerical calculations of $R_{s}$ for 
a conventional superconductor, indium. In the calculations, we take the realistic
parameters for indium: $\lambda$ = 0.81 (Ref.~\cite{Carb}), $\Delta_{0} (T) = 1.9T_{c} \tanh
[1.81(T_{c}/T-1)^{1/2}]$ (Ref.~\cite{Carb}), and $\theta_{D}$ = 111 K
(Ref.~\cite{Chan}).
We use three simple gap functions to characterize gap anisotropy: $\Delta
(\vec{\Omega})$ = 1 for isotropic $s$-wave gap ($s$-wave-1); $\Delta
(\vec{\Omega}) = (1 - 0.02\cos 4\phi)$ for a slightly anisotropic $s$-wave
gap ($s$-wave-2); $\Delta
(\vec{\Omega}) = (1 - 0.9\cos 4\phi)$ for a highly anisotropic $s$-wave
gap ($s$-wave-3). Fig.~1 shows numerically calculated results of
$R_{s}$ for the
three anisotropic $s$-wave gaps together with the measured $R_{s}$ for pure
indium. It is remarkable that the calculated curve for the isotopic $s$-wave gap
($s$-wave-1)
is slightly off from the experimental data while the curve for
the slightly anisotropic $s$-wave gap ($s$-wave-2) almost 
coincides with the data. In contrast, 
when the gap becomes highly anisotropic (e.g., $s$-wave-3), the Hebel-Slichter peak
is completely removed. This suggests that a highly anisotropic $s$-wave 
gap can completely suppress the coherenec peak. The numerical results for indium 
thus provide important insight 
into how sensitively the coherence peak changes with the gap anisotropy.

\begin{figure}[htb]
     \vspace{-0.3cm}
    \includegraphics[height=12cm]{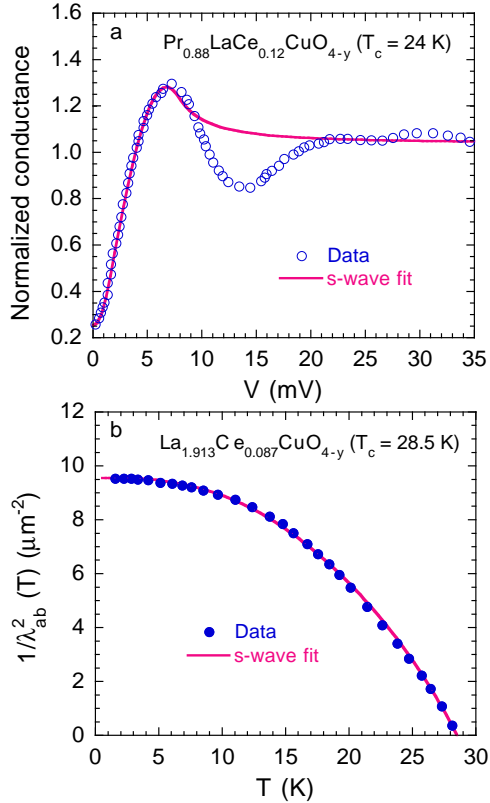}
     \vspace{-0.3cm}
 \caption[~]{a) Normalized tunneling conductance of a
electron-doped Pr$_{0.88}$LaCe$_{0.12}$CuO$_{4-y}$ crystal ($T_{c}$ = 24
K). The solid line is
numerically calculated curve using an anisotropic $s$-wave gap function: $\Delta_{1}(0)$ =
$4.9(|1.43\cos 2\theta - 0.43\cos 6\theta|  + 0.2)$ meV. b) Temperature dependence of $1/\lambda_{ab}^{2}$ for 
La$_{1.913}$Ce$_{0.087}$CuO$_{4-y}$ ($T_{c}$ = 28.5 K).  The solid line
is the
numerically calculated curve using an anisotropic $s$-wave gap function:
$\Delta_{1}(0)$ =
$3.5(|1.43\cos 2\theta - 0.43\cos 6\theta | + 0.35)$ meV. } 
\end{figure}

In order to address whether the absence of the coherence peak in the
$R_{s}$ data of Pr$_{0.91}$LaCe$_{0.09}$CuO$_{4-y}$ can be also
explained in terms of a highly anisotropic $s$-wave gap, we need
to extract the gap functions from independent experimental results such as
tunneling spectra and the in-plane magnetic penetration depth
$\lambda_{ab}(T)$. Fig.~2a shows
normalized tunneling conductance of a
electron-doped Pr$_{0.88}$LaCe$_{0.12}$CuO$_{4-y}$ crystal ($T_{c}$ = 24
K). The normalized tunneling spectrum is reproduced from Ref.~\cite{Zhao09}. The spectrum was taken on the top CuO$_{2}$ plane using
scanning tunneling microscopy (STM) \cite{Nie}. We can numerically calculate the
tunneling conductance using the following equation \cite{Suzuki}:

\begin{equation}
\frac{dI}{dV} \propto \int_{0}^{2\pi}p(\theta 
-\theta_{0}) Re[\frac{eV - i\Gamma}{\sqrt{(eV - i\Gamma)^{2} - 
\Delta_{1}^{2}(\theta)}}] d\theta,
\end{equation}
where $\theta$ is the angle measured 
from the Cu-O bonding direction, $\Gamma$ is the life-time broadening parameter of an
electron, and $p(\theta -\theta_{0})$ is the angle dependence of the 
tunneling probability and equal to $\exp [-\beta \sin^{2}(\theta -\theta_{0})]$. 
The solid line is the
numerically calculated curve using $\Gamma$ = 0.70 meV, $\beta$ = 5.3, $\theta_{0}$ =
$\pi/4$, and  an anisotropic $s$-wave gap function: $\Delta_{1}(0)$ =
$4.9(|1.43\cos 2\theta - 0.43\cos 6\theta | + 0.2)$ meV.
The finite life-time broadening parameter $\Gamma$ of an electron may be caused
by disorder and inhomogeneities. This life-time broadening effect at
zero temperature should be also
incorporated into the above expressions for $R_{s}$ by replacing
$\omega$ with $\omega-i\Gamma$.

Figure 2b shows the temperature dependence of $1/\lambda_{ab}^{2}$ for 
La$_{1.913}$Ce$_{0.087}$CuO$_{4-y}$ ($T_{c}$ = 28.5 K), which 
has a similar doping level as Pr$_{0.91}$LaCe$_{0.09}$CuO$_{4-y}$. The
data are digitized from Ref.~\cite{Skin}. We can 
numerically calculate 
the temperature dependence of 
$\lambda_{ab}^{2}(0)/\lambda_{ab}^{2}(T)$ for an anisotropic gap function using 
the following equation: \cite{Jacobs}
\begin{equation}
\frac{\lambda_{ab}^{2}(0)}{\lambda_{ab}^{2}(T)} = 1 + 
(1/\pi)\int_{0}^{2\pi}\int_{0}^{\infty}d\theta d\epsilon 
\frac{\partial f}{\partial E}.
\end{equation}
Here $E = \sqrt{\epsilon^{2}+ \Delta_{1}^{2}(\theta,T)}$. The solid line is the
numerically calculated curve using an anisotropic $s$-wave gap function:
$\Delta_{1}(T, \theta)$ =
$3.5\tanh [1.81(T_{c}/T-1)^{1/2}](|1.43\cos 2\theta - 0.43\cos 6\theta | + 0.35)$ meV. 
If we assume that the 
superconducting gap is proportional to $T_{c}$, 
then the bulk superconducting gap of Pr$_{0.91}$LaCe$_{0.09}$CuO$_{4-y}$ inferred from
the penertation depth data is 2.95$(|1.43\cos 2\theta - 0.43\cos 6\theta | + 0.35)$
meV.

Now we can use the inferred gap functions from both STM and magnetic
penetration depth to calculate $R_{s}$ for
Pr$_{0.91}$LaCe$_{0.09}$CuO$_{4-y}$. In the calculations, we take the realistic parameters
for the electron-doped cuprate: $\lambda$ = 1.0 (Ref.~\cite{Huang}) and $\theta_{D}$ =
384 K (Ref.~\cite{Ham}). Since the $R_{s}$ data
were taken in a magnetic field of 6.2 T which suppresses $T_{c}$ to
about 20 K, we scale down the gap size proportional to $T_{c}$. For
the gap function extracted from the tunneling spectrum, the same
life-time broadening parameter ($\Gamma$ = 0.7 meV) in the tunneling
spectrum is used to 
calculate $R_{s}$. This
parameter only influences $R_{s}$ at low temperatures and has little 
effect on $R_{s}$ close to $T_{c}$. 

\begin{figure}[htb]
     \vspace{-0.2cm}
    \includegraphics[height=6.0cm]{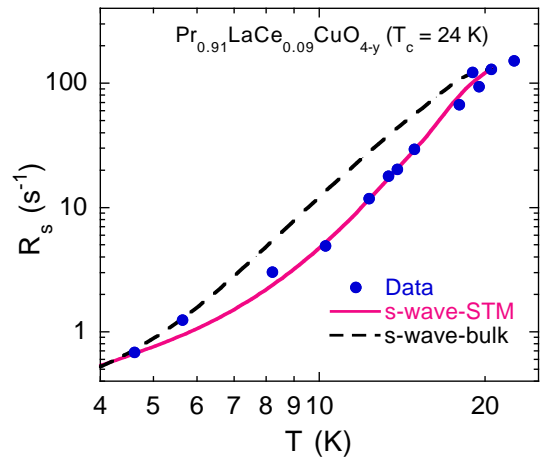}
     \vspace{-0.3cm}
 \caption[~]{Nuclear magnetic spin-lattice relaxation rate ($R_{s}$)
 in Pr$_{0.91}$LaCe$_{0.09}$CuO$_{4-y}$ (solid circles). The solid and
 dashed lines are the numerically calculated $R_{s}$ 
 curves for
the $s$-wave gap functions inferred from STM (denoted as STM) and from
the magnetic penetration 
depth (denoted as bulk). } 
\end{figure}

The calculated results are shown in Fig.~3. It is apparent that for
the gap function extracted from STM, the calculated $R_{s}$ curve
(solid line) agrees excellently
with the experimental data. On the other hand, for
the bulk gap function, the
calculated curve (dashed line) deviates significantly from the 
data. In all the cases, the coherence peaks are absent. Therefore, the absence of the
coherence peak in the $R_{s}$ data could be consistent with highly anisotropic $s$-wave
gaps.

Although the $R_{s}$ data can be excellently explained in
terms of the gap function inferred from STM, it does not necessarily imply that
the gap function deduced from STM is more realistic than that
inferred from the magnetic penetration depth. As a matter of fact, 
the above calculations do not take into account strong
electron-electron correlation. It was shown that strong electron-electron correlation
can reduce the coherence peak of $R_{s}$ for an $s$-wave gap and cause 
$R_{s}$ to drop more rapidly just below $T_{c}$ \cite{Sca}. For
the slightly underdoped Pr$_{0.91}$LaCe$_{0.09}$CuO$_{4-y}$,  the measured value 
of $T_{1}TK_{s}^{2}$ 
(where $T_{1}$ is the spin-lattice
relaxation time and $K_{s}$ is the spin part of Knight shift) is
a factor of 50 smaller than the expected value of noninteracting
electrons \cite{Zheng}. This implies strong electron-electron correlation. Therefore,
it is very likely that the $R_{s}$ data can be quantitatively explained in terms
of the bulk anisotropic $s$-wave gap if the strong electron-electron correlation
is taken into account.

\begin{figure}[htb]
    \vspace{-0.2cm}
    \includegraphics[height=6.0cm]{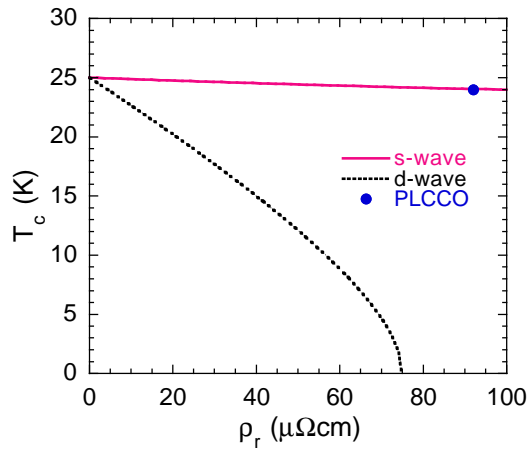}
     \vspace{-0.3cm}
 \caption[~]{The calculated curve of $T_{c}$ versus residual resistivity 
 in terms of any $d$-wave gap (dotted line) and the bulk anisotropic gap (solid line): $\Delta$ = 2.95$(|1.43\cos 2\theta - 0.43\cos 6\theta | + 0.35)$
meV. In the calculations, we use $\hbar\Omega_{p}^{*}$ =
0.62~eV. The solid circle is a data point for
 Pr$_{0.91}$LaCe$_{0.09}$CuO$_{4-y}$ (PLCCO) \cite{Zheng}.
 }
\end{figure}

In order to unambiguously distinguish between any $d$-wave and anisotropic
$s$-wave gap symmetries, we study the response of a superconductor to 
nonmagnetic impurities or disorder. 
The nonmagnetic impurity pair-breaking effect is both bulk- and
phase-sensitive. 
This is because the rate of $T_{c}$ suppression by nonmagnetic impurities \cite{Op} is determined
by the value of the Fermi surface (FS) average $<\Delta
(\vec{k})>_{FS}$, which depends sensitively on the phase of the gap
function. More specifically, the rate is proportional to 
a parameter $\chi$ = $1-(<\Delta (\vec{k})>_{FS})^{2}/<\Delta^{2}(\vec{k})>_{FS}$.
It is easy to show that $\chi$ = 1 for any $d$-wave gap and $\chi$
= 0.058 for the highly anisotropic $s$-wave gap: $\Delta$ = 2.95$(|1.43\cos 2\theta - 0.43\cos 6\theta | + 0.35)$
meV.  An equation to describe the pair-breaking effect by
nonmagnetic impurities (or defects) is given by \cite{Op}
\begin{equation}\label{pairbreak}
\ln \frac{T_{c0}}{T_{c}}= \chi[\Psi (\frac{1}{2} + \frac{0.122
	  (\hbar\Omega_{p}^*)^{2}\rho_{r}}{T_{c}}) - \Psi
	  (\frac{1}{2})],
\end{equation}	  
where $\hbar\Omega_{p}^{*}$ is the renormalized plasma energy
\cite{Op,Rad} in units
of eV, $\rho_{r}$ is the residual resistivity in units of
$\mu\Omega$cm, and $\Psi$ is the digamma function. The lower limit of 
$\hbar\Omega_{p}^{*}$ is equal to $\hbar\Omega_{s}^{*}$ which is
related to the zero-temperature
penetration depth $\lambda_{ab}(0)$. For
La$_{1.913}$Ce$_{0.087}$CuO$_{4-y}$, $\lambda_{ab}(0)$
= 320 nm (Ref.~\cite{Skin}), leading to $\hbar\Omega_{s}^{*}$ = 0.62 eV. Figure 4 shows the calculated 
curves of $T_{c}$ versus residual resistivity in terms of 
any $d$-wave gap (dotted line) and the bulk anisotropic $s$-wave gap (solid line): $\Delta$ = 2.95$(|1.43\cos 2\theta - 0.43\cos 6\theta | + 0.35)$
meV. For the $d$-wave gap, $T_{c}$ is suppressed to 0 at $\rho_{r}$
= 74.6~$\mu\Omega$cm when the lower limit of $\hbar\Omega_{p}^{*}$ =
0.62~eV is used. In Pr$_{1.85}$Ce$_{0.15}$CuO$_{4-y}$ ($T_{c}$ = 20 K), $\hbar\Omega_{p}^{*}$
is found to be a factor of 1.65 larger than $\hbar\Omega_{s}^{*}$ due to a 
finite mean-free path \cite{Homes}. If we increase $\hbar\Omega_{p}^{*}$ 
of Pr$_{0.91}$LaCe$_{0.09}$CuO$_{4-y}$ by the same factor (1.65), $T_{c}$ will be suppressed to 0 at $\rho_{r}$
= 27.4~$\mu\Omega$cm for the $d$-wave gap. The measured large $\rho_{r}$
of 92 $\mu\Omega$cm and nearly
optimal  $T_{c}$ of 24 K in  Pr$_{0.91}$LaCe$_{0.09}$CuO$_{4-y}$
rules out any $d$-wave gap symmetry.

In summary, the absence of the Hebel-Slichter
peak in the nuclear spin-lattice relaxation rate of Pr$_{0.91}$LaCe$_{0.09}$CuO$_{4-y}$
can be well explained by a highly anisotropic $s$-wave gaps inferred respectively from the magnetic penetration depth and 
scanning tunneling microscopy.  
The observed weak nonmagnetic pair-breaking effect provides unambiguous
evidence for a highly anisotropic $s$-wave gap in this slightly underdoped $n$-type cuprate.

\bibliographystyle{prsty}

\end{document}